\documentclass[
    ,final            
  ]
  {aipproc}

\layoutstyle{6x9}
\def\Pom{{\bf I\!P}}

\begin{document}

\title{Diffractive production of charm quark/antiquark pairs\\
at RHIC and LHC}


\author{Marta {\L}uszczak}{
  address={University of Rzesz\'ow, PL-35-959 Rzesz\'ow, Poland}
}


\author{Antoni Szczurek}{
  altaddress={University of Rzesz\'ow, PL-35-959 Rzesz\'ow, Poland},
  address={Institute of Nuclear Physics PAN, PL-31-342 Cracow, Poland}
}


\begin{abstract}

We have discussed single and central diffractive production
of $c \bar c$ pairs in the Ingelman-Schlein model. In these
calculations we have included diffractive parton distributions
obtained by the H1 collaboration at HERA and absorption effects
neglected in some early calculations in the literature.
The absorption effects which are responsible for the 
naive Regge factorization breaking cause that the cross section 
for diffractive processes is much smaller than that for the fully 
inclusive case, but could be measured at RHIC and LHC by 
imposing special condition on rapidity gaps.
We discuss also different approaches to diffractive production
of heavy quark/antiquark  \cite{Alves,Yuan,Kopeliovich}.
The particular mechanism is similar to the diffractive dissociation
of virtual photons into quarks, which drives diffractive deep inelastic
production of charm in the low-mass diffraction, or large $\beta$-region.

\end{abstract}

\maketitle


\section{Introduction}
\label{intro}

In this presentation we discuss diffractive processes (single and central)
in the framework of Ingelman-Schlein model corrected for absorption.
The formalism and more details has been shown and discussed
elsewhere \cite{LMS2011}.
Such a model was used in the estimation of several diffractive
processes \cite{diffractive_dijets,double_pomeron,double_diff,H1}.
The absorption corrections are necessary to understand a huge 
Regge-factorization breaking observed in single and central 
production at Tevatron.
We mention also the QCD mechanism of diffractive dissociation of gluons into
heavy quark pairs. However, the associated formalism and our results (in
progress) will be discussed elsewhere \cite{LSS2012}.

\section{A sketch of formalism}
\label{formalism}

\begin{figure}[!ht]
a)  \includegraphics[height=.16\textheight]{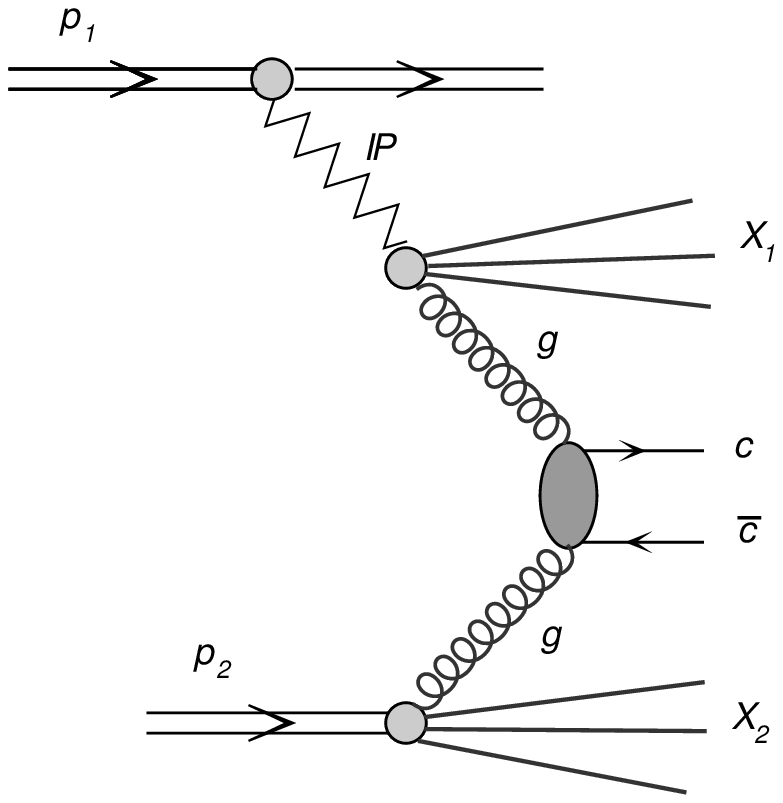}
b)  \includegraphics[height=.16\textheight]{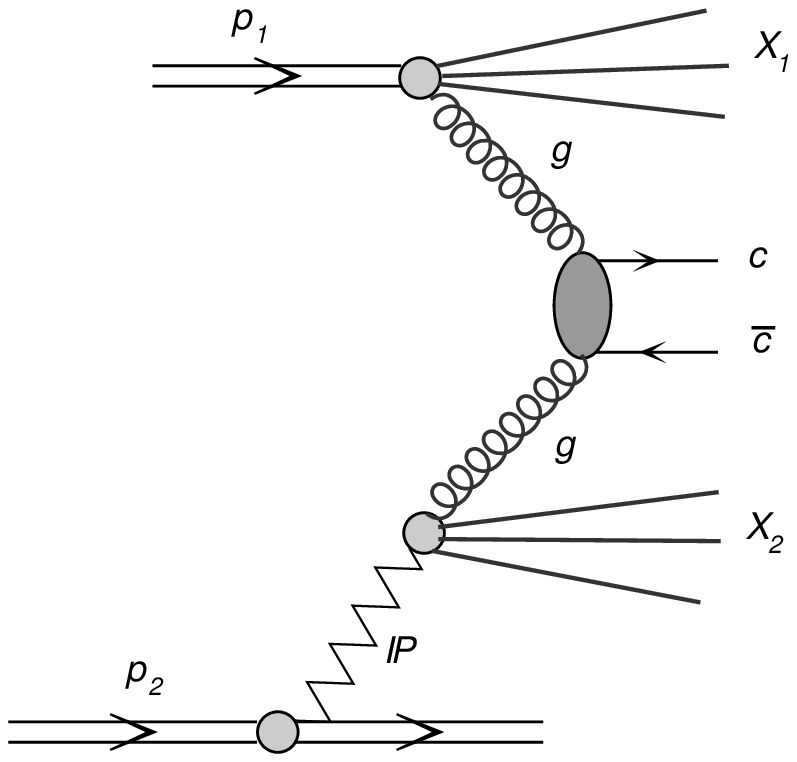}
\caption{
The mechanism of single-diffractive production of $c \bar c$.
}
\label{fig:1}
\end{figure}

\begin{figure}[!ht]
a)  \includegraphics[height=.16\textheight]{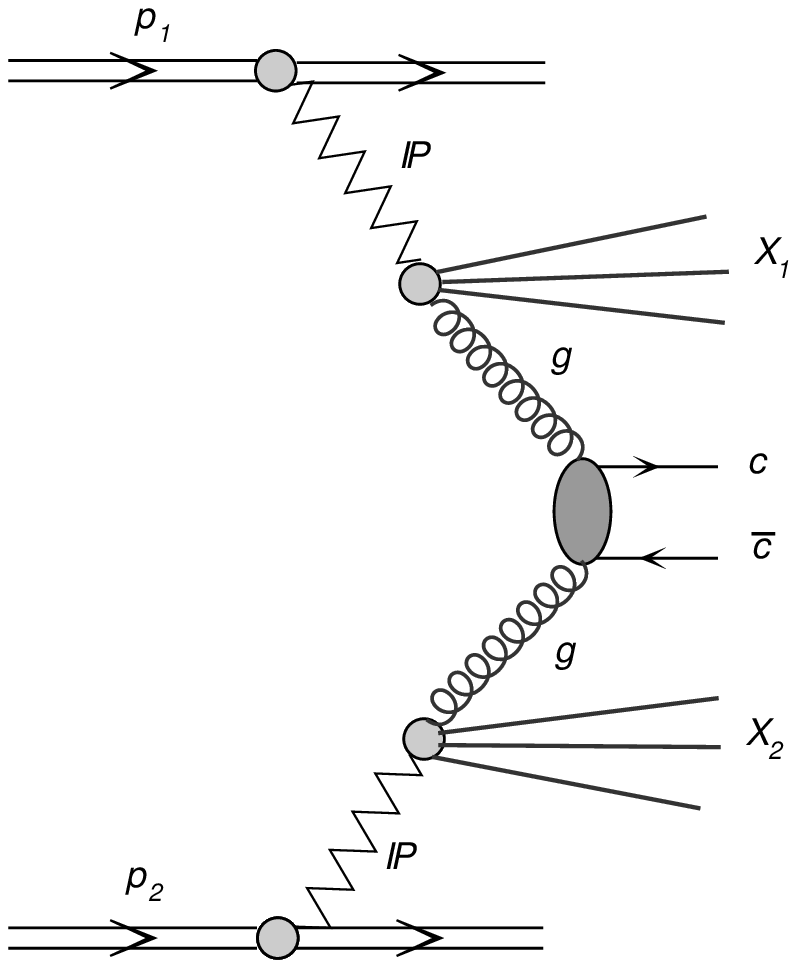}
\caption{
The mechanism of central-diffractive production of dileptons. 
}
\label{fig:2}
\end{figure}

The mechanisms of the diffractive production 
of heavy quarks ($c \bar c$) discussed here are shown in 
Figs.\ref{fig:1},\ref{fig:2}.

In the following we apply the Ingelman and Schlein approach. 
In this approach one assumes that the Pomeron has a
well defined partonic structure, and that the hard process
takes place in a Pomeron--proton or proton--Pomeron (single diffraction) 
or Pomeron--Pomeron (central diffraction) processes.
In this approach corresponding differential cross sections can be
written as
\begin{eqnarray}
{d \sigma_{SD} \over dy_{1} dy_{2} dp_{t}^2} =  K {\Big| M \Big|^2 \over 16 \pi^2 \hat{s}^2} 
\,\Big [\, \Big( x_1 q_f^D(x_1,\mu^2) 
\, x_2 \bar q_f(x_2,\mu^2) \Big) \, 
+ \Big( x_1 \bar q_f^D(x_1,\mu^2)
\, x_2  q_f(x_2,\mu^2) \Big) \, \Big ] ,
\nonumber \\ 
\label{SD}
\end{eqnarray}
\begin{eqnarray}
{d \sigma_{CD} \over dy_{1} dy_{2} dp_{t}^2} =  K {\Big| M \Big|^2 \over 16 \pi^2 \hat{s}^2} 
\,\Big [\, \Big( x_1 q_f^D(x_1,\mu^2) 
\, x_2 \bar q_f^D(x_2,\mu^2) \Big) \, 
+ \Big( x_1 \bar q_f^D(x_1,\mu^2)
\, x_2  q_f^D(x_2,\mu^2) \Big) \,\Big ] 
\nonumber \\ 
\label{DD}
\end{eqnarray}
for single-diffractive and central-diffractive production, 
respectively.

The 'diffractive' quark distribution of
flavour $f$ can be obtained by a convolution of the flux of Pomerons
$f_\Pom(x_\Pom)$ and the parton distribution in the Pomeron 
$q_{f/\Pom}(\beta, \mu^2)$:
\begin{eqnarray}
q_f^D(x,\mu^2) = \int d x_\Pom d\beta \, \delta(x-x_\Pom \beta) 
q_{f/\Pom} (\beta,\mu^2) \, f_\Pom(x_\Pom) \, 
= \int_x^1 {d x_\Pom \over x_\Pom} \, f_\Pom(x_\Pom)  
q_{f/\Pom}({x \over x_\Pom}, \mu^2) \, . \nonumber \\
\end{eqnarray}
The flux of Pomerons $f_\Pom(x_\Pom)$ enters in the form integrated over 
four--momentum transfer 
\begin{eqnarray}
f_\Pom(x_\Pom) = \int_{t_{min}}^{t_{max}} dt \, f(x_\Pom,t) \, ,
\label{flux_of_Pom}
\end{eqnarray}
with $t_{min}, t_{max}$ being kinematic boundaries.

Both pomeron flux factors $f_{\Pom}(x_{\Pom},t)$ as well 
as quark/antiquark distributions in the pomeron were taken from 
the H1 collaboration analysis of diffractive structure function
and diffractive dijets at HERA \cite{H1}. 
The factorization scale for diffractive parton distributions is taken as
$\mu^2 = \hat s$.

\begin{figure}[!ht]
  \includegraphics[height=.17\textheight]{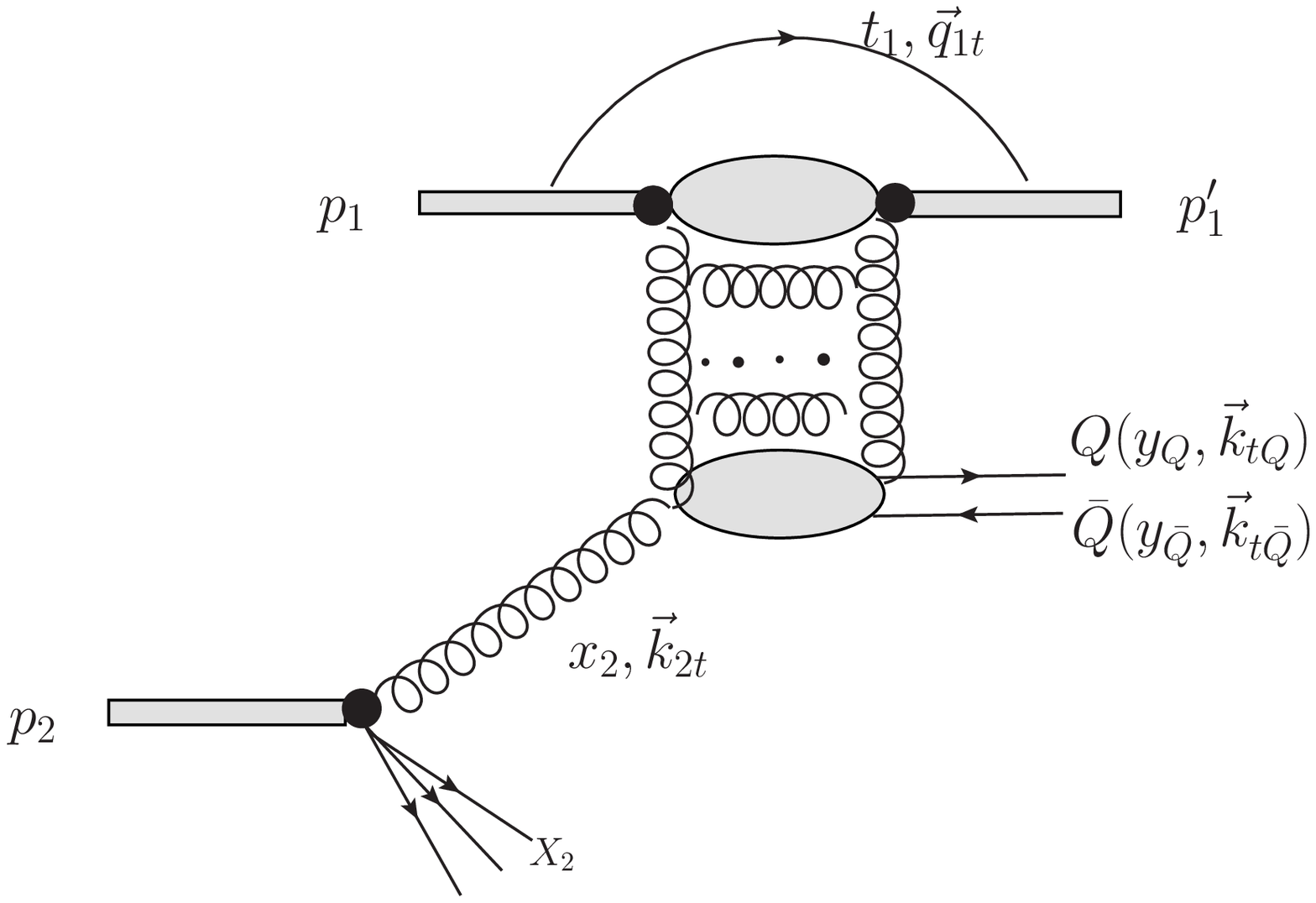}
  \includegraphics[height=.22\textheight]{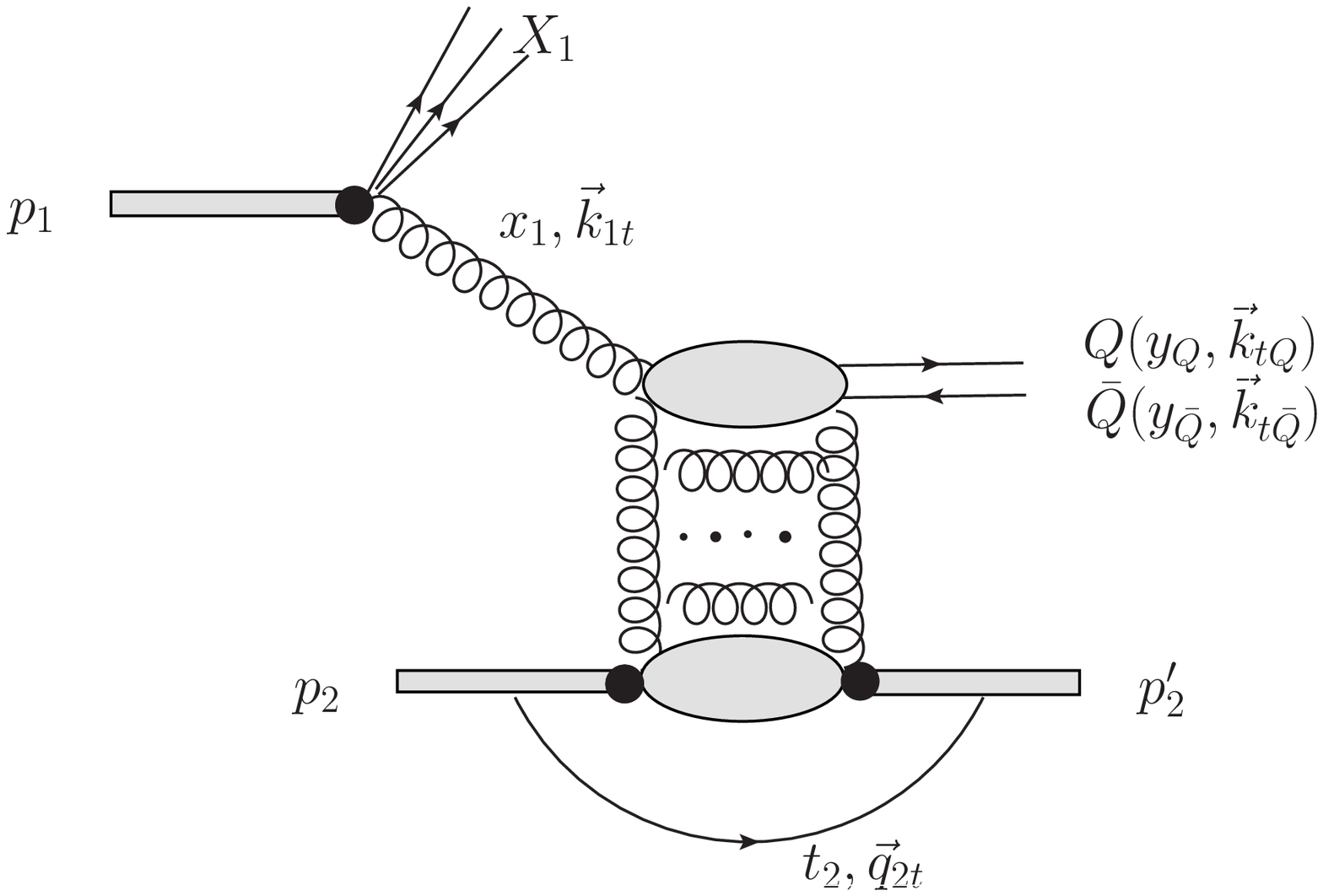}
\caption{
}
\label{fig:3}
\end{figure}

In Ref.\cite{LSS2012} we shall study in detail mechanism of diffractive 
dissociation of gluons into heavy quark pairs presented in Fig.\ref{fig:3}.
There we shall present relevant amplitudes, calculate relevant
differential cross sections and compare to the Ingelman-Schlein model 
discussed here.

\section{Results}
\label{results}

\begin{figure}[!ht]
  \includegraphics[height=.23\textheight]{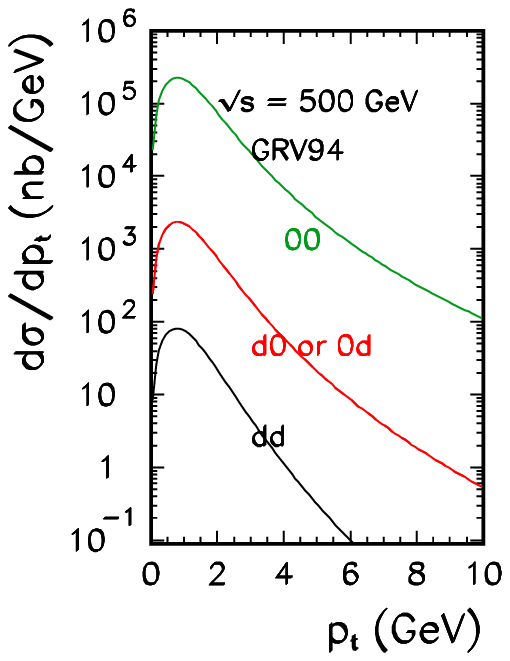}
  \includegraphics[height=.23\textheight]{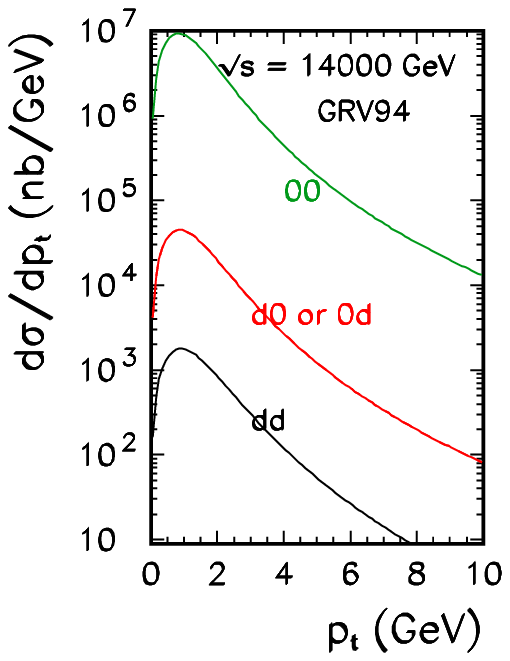}
\caption{Transverse momentum distribution of $c$ quarks (antiquarks)
for RHIC energy $\sqrt{s} =$ 500 GeV (left panel) and for LHC energy
$\sqrt{s}$ = 14 TeV (right panel) for the GRV94 gluon distributions.
The result for single diffractive (0d or d0), central diffractive (dd) 
mechanisms are compared with that for the standard gluon-gluon fusion (00).
}
\label{fig:diff_dsig_dpt_diff}
\end{figure}

In Fig.\ref{fig:diff_dsig_dpt_diff} we show transverse momentum
distributions of charm quarks (or antiquarks). The distribution
for single diffractive component is smaller than that for the
inclusive gluon-gluon fusion by almost 2 orders of magnitude. Our
results include gap survival factor \cite{LMS2011}. 
The cross section for central diffractive component is smaller by 
additional order of magnitude.


\begin{figure}[!h]
 \includegraphics[height=.23\textheight]{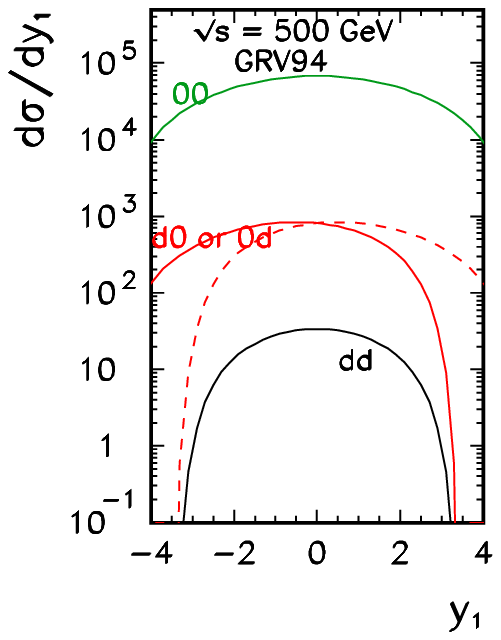}
  \includegraphics[height=.23\textheight]{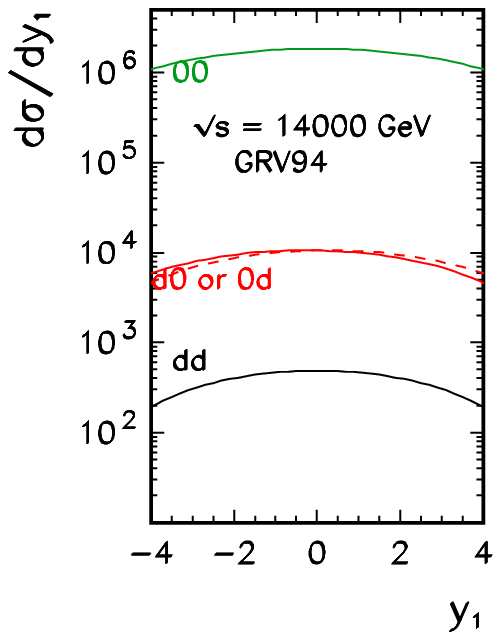}
\caption{Rapidity distribution of $c$ quarks (antiquarks)
for RHIC energy $\sqrt{s} =$ 500 GeV (left panel) and 
for LHC energy $\sqrt{s}$ = 14 TeV (right panel) for the GRV94 gluon 
distributions. The result for single diffractive (0d or d0), 
central diffractive (dd) mechanisms
are compared with that for the standard gluon-gluon fusion (00).
}
\label{fig:diff_dsig_dy1_diff}
\end{figure}

In Fig.\ref{fig:diff_dsig_dy1_diff} we show distributions in quark
(antiquark) rapidity. We show separately contributions of two different 
single-diffractive components (which give the same distributions in
transverse momentum) and the contribution of central-diffractive
component in Fig.\ref{fig:diff_dsig_dpt_diff}.
When added together the single-diffractive components produce a 
distribution in rapidity similar in shape to that for the standard 
inclusive case.


\section{Conclusions}

The cross section for single and central diffraction is rather small
compared to the dominant gluon-gluon fusion component.
However, a very specific final state (smaller multiplicity, forward
protons) should allow its identification
by imposing special conditions on the one-side (single-diffractive
process) and both-side (central diffractive process) rapidity gaps.
We hope that such an analysis is possible at LHC. Special care
should be devoted to the observation of the exclusive $c \bar c$
production, i.e. the process $p p \to p p c \bar c$.
Without a special analysis of the final state multiplicity
the exclusive $c \bar c$ production may look like an inclusive
central diffraction \cite{LMS2011}.
We have started detailed analysis of diffractive dissociation 
of gluons into $c \bar c$ pairs as a competetive process
to the one discussed here.





\bibliographystyle{aipproc}   

\bibliography{sample}

\IfFileExists{\jobname.bbl}{}
 {\typeout{}
  \typeout{******************************************}
  \typeout{** Please run "bibtex \jobname" to optain}
  \typeout{** the bibliography and then re-run LaTeX}
  \typeout{** twice to fix the references!}
  \typeout{******************************************}
  \typeout{}
 }


\end{document}